\documentclass{article}

\usepackage{PRIMEarxiv}

\usepackage[utf8]{inputenc} 
\usepackage[T1]{fontenc}    
\usepackage{hyperref}       
\usepackage{url}            
\usepackage{booktabs}       
\usepackage{amsfonts}       
\usepackage{nicefrac}       
\usepackage{microtype}      
\usepackage{lipsum}
\usepackage{fancyhdr}       
\usepackage{graphicx}       
\graphicspath{{media/}}     
\usepackage{placeins}
\usepackage{subcaption}
\title{Terahertz Spatio-Temporal Deep Learning Computed Tomography
}

\author{
  Yi-Chun Hung \\
  Department of Electrical and Computer Engineering \\ University of California, Los Angeles \\
  \texttt{yichunhung@g.ucla.edu} \\
  \And
  Ta-Hsuan Chao \\
  Department of Electrical Engineering \\
  Stanford University \\
  \texttt{thchao@stanford.edu} \\
  \AND
  Pojen Yu\\
  Department of Electrical Engineering \\
  National Tsing Hua University, Hsinchu \\
  \texttt{jerry321ab@gmail.com}
  \AND
  Shang-Hua Yang \\
  Department of Electrical Engineering \\
  National Tsing Hua University, Hsinchu \\
  \texttt{shanghua@ee.nthu.edu.tw}
}

\begin{document}
\maketitle

\begin{abstract}
Terahertz computed tomography (THz CT) has drawn significant attention because of its unique capability to bring multi-dimensional object information from invisible to visible. However, current physics-model-based THz CT modalities present low data use efficiency on time-resolved THz signals and low model fusion extensibility, limiting their application fields' practical use. In this paper, we propose a supervised THz deep learning computed tomography (THz DL-CT) framework based on time-domain information. THz DL-CT restores superior THz tomographic images of 3D objects by extracting features from spatio-temporal THz signals without any prior material information. Compared with conventional and machine learning based methods, THz DL-CT delivers at least 50.2\%, and 52.6\% superior in root mean square error (RMSE) and structural similarity index (SSIM), respectively. Additionally, we have experimentally demonstrated that the pretrained THz DL-CT model can generalize to reconstruct multi-material systems with no prerequisite information. THz CT through the DL data fusion approach provides a new pathway for non-invasive functional imaging in object investigation.
\end{abstract}


\section{Introduction}
Thanks to the rapid advancement of terahertz (THz) systems, nowadays, a great variety of THz technologies -- including spectroscopy \cite{beard2002terahertz_TDS_properties_and_applications, dexheimer2017terahertz_TDS_principles_applications}, near-field imaging \cite{serita2012scanning_near_field_imaging}, hyperspectral imaging \cite{kawase2003non}, time-resolved imaging \cite{zhong2005nondestructive}, or the combination of the abovementioned \cite{wu2019sub} -- have no more leashed under laboratory environment and become powerful engines that deliver unique functionalities in the real-world settings \cite{xie2014application_protein, shen2008development_pharmaceutical_tablet}. As THz wave can transmit through many optically opaque materials carrying the material fingerprints along with their geometric information, the THz system has been well-launched in many application fields, such as chemical identification, security screening, defect inspection, pharmaceutical quality control, remote sensing, and imaging \cite{mittleman1996t, mittleman1999recent, jansen2010terahertz, zhang2003terahertz_carrier_dynamics}. Among those applications, remote sensing and non-destructive imaging are extremely important due to their efficacy in revealing material properties and the geometric structure of the object under test. Among different types of THz systems, THz time-domain spectroscopy (THz-TDS) systems are commonly used to digitally dissect inner object slices through the time-resolved THz electric field signals \cite{zhong2005nondestructive, shen20053}, which encode rich object information from ultrafast molecular dynamics, dielectric responses, electrical properties to conformational behaviors \cite{ulbricht2011carrier_dynamics, padilla2006dynamical_metameterial}. As the THz time-domain signal records the light-matter interaction behaviors along the traveling path, the THz temporal signal features, including peak THz electric field strength, time-delay, echo, and polarity changes, play essential roles to reconstruct object information. Apart from providing time-domain information, the THz-TDS system can provide a broadband spectral response in amplitude, and phase \cite{davies2008terahertz, petrov2016application}. Based on the multi-dimensional data cube, recent advancements of THz-TDS imaging modalities have successfully demonstrated a great variety of physical parameter extraction with deep-subwavelength spatial resolution levels \cite{huber2008terahertz, shin2018qualitative}. Nowadays, the data acquisition time and signal-to-noise ratio of THz-TDS systems have achieved orders of magnitude enhancement than its first invention \cite{yardimci2021broadband, elzinga1987pump}. It widely opens the scope of non-destructive evaluation in material inspection, spatial dimension, and functional 3D mapping of different types of optically opaque objects.

Many works aim to extract different physical characteristics of tested objects by analyzing signal features from the retrieved raw data. For example, time delay and phase spectrum are highly relative to the thickness of objects \cite{duvillaret1999highly}; field strength and power spectrum are related to absorption coefficients, electrical conductivity, doping concentration, and chemical maps inside the objects \cite{davies2008terahertz, huber2008terahertz}. Zhong et al. \cite{zhong2005nondestructive} used time delay information to reveal voids and delamination defects in the object; Shen et al. \cite{shen20053} analyzed the subsurface structure and the chemical mapping of the multilayered object based on peak delay information. However, those early works still presented a certain degree of inaccuracy in capturing time delay due to peak broadening and signal distortion. To further improve the accuracy of peak delay estimation, Takayanagi et al. \cite{takayanagi2009high} combined the information with Wiener filters and deconvolution methods to resolve a 2 $\mu$m-thick GaAs layer, which reaches one-hundredth of the THz wavelengths. In \cite{stubling2019application}, the sparse deconvolution algorithm is performed to reveal the 3D structure of an artificially mummified ancient Egyptian human left hand. Other than design filters and performing deconvolution, Wu et al. \cite{wu2019sub} characterized the THz temporal profiles in object interface regions and proposed the differential pulse delay method to deblur THz images. With the utilization of time-domain information, especially the time delay and elongated temporal features, the geometric details of 3D objects can be further accurately reconstructed.

Although the THz-TDS modality is well-suited for non-destructive 3D imaging, the ambient environment conditions would still considerably impact time-resolved THz signal quality, especially the water vapor and additive surrounding noise levels. It causes undesired lower signal-to-noise ratio (SNR) and peak broadening issues, leading to the inaccuracy of reconstructed 2D THz images. Recur et al. \cite{recur2012propagation} developed a THz spatial beam propagation model instead of the conventional ray-tracing model to increase the reconstructed 3D image quality. In \cite{recur2014ordered}, the same research group further developed a statistic approach with THz beam properties to achieve images with sub-millimeter spatial resolution. Additionally, features in the frequency domain are also adopted to restore images. Png et al. \cite{png2005terahertz} extracted the phase information of THz waves to observe the thickness mapping of a leaf. Compared with time delay information, spectroscopic information provides superior performance in thickness estimation. Most importantly, it reveals the material fingerprints and composition along the THz pathway. Kawase et al. \cite{kawase2003non} analyze the attenuated THz power of selective frequencies to demonstrate illicit drug separation with high image contrast due to the distinctive light-matter interaction among frequencies. Singh et al. \cite{singh2020three} estimates the 3D water mapping of succulent Agave victoriae-reginae leaves by fitting the dielectric function of succulent Agave victoriae-reginae leaves, which is retrieved by the THz measured spectrum to the well-studied dielectric functions of water, solid/dry tissue, and air. Among most THz imaging modalities, the prior knowledge of material selections and geometric configurations plays a pivotal role in bridging the THz signal and functional mappings of tested objects. Most of the time, this prerequisite information cannot be obtained in the real world, which severely constrains the applicable scenarios of THz-TDS systems. In this regard, our motivation is to develop a universal THz imaging framework without prior knowledge of tested objects and be capable of generalizing this framework to multi-material systems with arbitrary object arrangement configurations. 

We propose the THz deep learning computed tomography (THz DL-CT) framework, a supervised convolutional neural network without predefined features and prior information. Based on convolutional neural network (CNN) blocks \cite{lecun1998gradient}, THz DL-CT extracts informative features from both spatial and spectral domains. In this paper, we investigate the framework's performance and compare it with conventional and machine learning approaches. We also demonstrate that the extracted data-driven features are interpretable and be highly correlated to physic implication based on saliency maps \cite{zhou2016learning}, which are commonly used for visual attention in computer vision and artificial intelligence related fields.

\section{Experimental System}

\begin{figure}
    \centering
    \includegraphics[width=\linewidth]{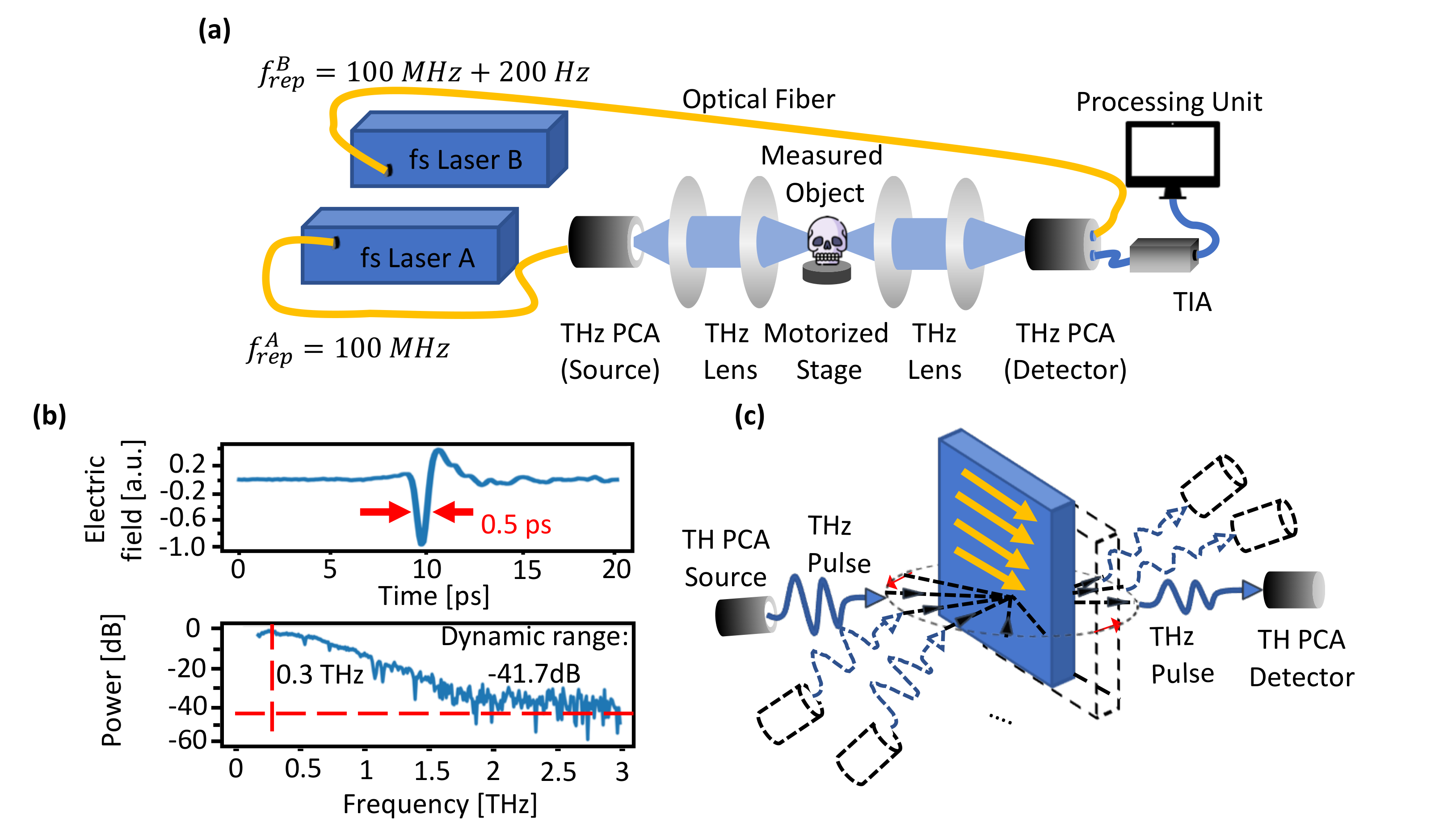}
     \caption{(a) Illustration of the asynchronous optical sampling (ASOPS) THz time-domain spectroscopy (THz-TDS) system. Two asynchronous femtosecond lasers are operated at 100 MHz and 100 MHz + 200 Hz. After THz waves travel through the measured object, THz waves are focused to a THz photoconductive antenna (PCA) detector and amplified by a transimpedance amplifier (TIA). The amplified signals are then sampled by a data acquisition card and processed by a computer. (b) The reference THz signal of the ASOPS THz-TDS system in time-domain and frequency domain. (c) The schematic diagram of data the ASOPS THz-TDS system scanning protocol. Measured objects are raster-scanned by a two-axis mechanical stage and a rotational stage with a step size of 0.25 mm and a 6 degree, respectively.}
     \label{fig:THzTDS}
\end{figure}

The experimental setup is based on a THz-TDS system that records the geometric information of the tested object in the time-resolved THz electric field signals. The temporal features of the THz signals (i.e., time delay, attenuated temporal profiles) influenced by the material dielectric response and object geometry would then use for THz computed tomography reconstruction. To speed up the data acquisition process, we use an asynchronous optical sampling (ASOPS) THz-TDS system (Menlo TERA ASOPS, MenloSystems, Germany) \cite{stoica2008wideband} without any mechanical moving part. Compared with conventional THz-TDS systems, the data acquisition time of each time-resolved THz trace can be further shrunk down to a millisecond timescale, which is beneficial for the large-volume THz database establishment.

As shown in Fig. \ref{fig:THzTDS}(a), the ASOPS THz-TDS system is composed of two asynchronous femtosecond Er-doped fiber lasers at a central wavelength of 1560 nm, a pair of THz photoconductive antenna (PCA) source and detector, four plane-convex TPX lens with 50 mm focal length, a linear and rotation motorized stage for object position adjustment, and a transimpedance amplifier (TIA). The repetition rate of the two asynchronous femtosecond lasers is 100 MHz and 100 MHz + 200 Hz, respectively. Due to the slight difference in the repetition rates, the asynchronous pump-probe approach sequentially profiles the time-resolved THz signals by accumulating the time delay between the two lasers in every repetition cycle. The data acquisition rate of the processing unit is 20 MHz, equivalently delivering 0.1 ps temporal resolution of THz signals and THz frequency bandwidth up to 5 THz. Based on the system configuration, the ASOPS THz-TDS system provides the THz pulse signal with a 0.5 ps at full width at half maximum (FWHM), a 41.7 dB dynamic range from 0.3 THz to 3 THz, and a 5 ms data acquisition time for a single THz signal trace (as shown in Fig. \ref{fig:THzTDS}(b)).
Additionally, each time-resolved THz signal trace contains 100K sampling points corresponding to 10 ns in the time domain. Considering the combination of 2D raster-scanning, multiple projection angles, and time-resolved information, the meta-data size can reach the terabyte (TB) level, which requires high computation power for data processing. To address the computational issue, we select the object-dependent THz signals with a 100 ps timeframe (1K sampling points) around the THz pulse signal – for further data training and processing. This 100 ps temporal data segment can accommodate the measurement of objects in the size of the several centimeters level. Fig. \ref{fig:THzTDS}(c) shows the schematic diagram of the ASOPS THz-TDS system scanning protocol. The rotational scanning range is 180 degrees with a step of 6 degrees. The horizontal scanning range is 72 mm with a step of 0.25 mm, which is approximately one-fifth of the THz beam diameter at FWHM. The vertical scanning step is 0.25 mm, and the scanning range varies from the height of measured objects. For both model training and performance evaluation purposes, eight 3D-printed objects are fabricated by a fused deposition modeling (FDM) 3D printer with sub-millimeter spatial resolution, which is in the subwavelength range of the pulsed THz waves. We choose high impact polystyrene (HIPS) as the 3D printing material to prevent severe material absorption in several inches thick objects. The absorption spectrum of HIPS is lower than $10~\textit{cm}^{-1}$ over 0.1 - 1.2 THz frequency range. The ground truth cross-sectional images for training the THz DL-CT framework is generated from the cross-sections of digital 3D object models. Corresponding to the scanning configuration, the axial, lateral, and vertical resolution of the ground truth images are all 0.25 mm. Additionally, the bottom of every printed HIPS object contains the same alignment mark pattern. The alignment mark under the objects provides the sub-degree rotational alignment precision between ground truth images and the measured THz pulse signals.

\section{Dataset Description}

\begin{figure}
    \centering
    \includegraphics[width=\linewidth]{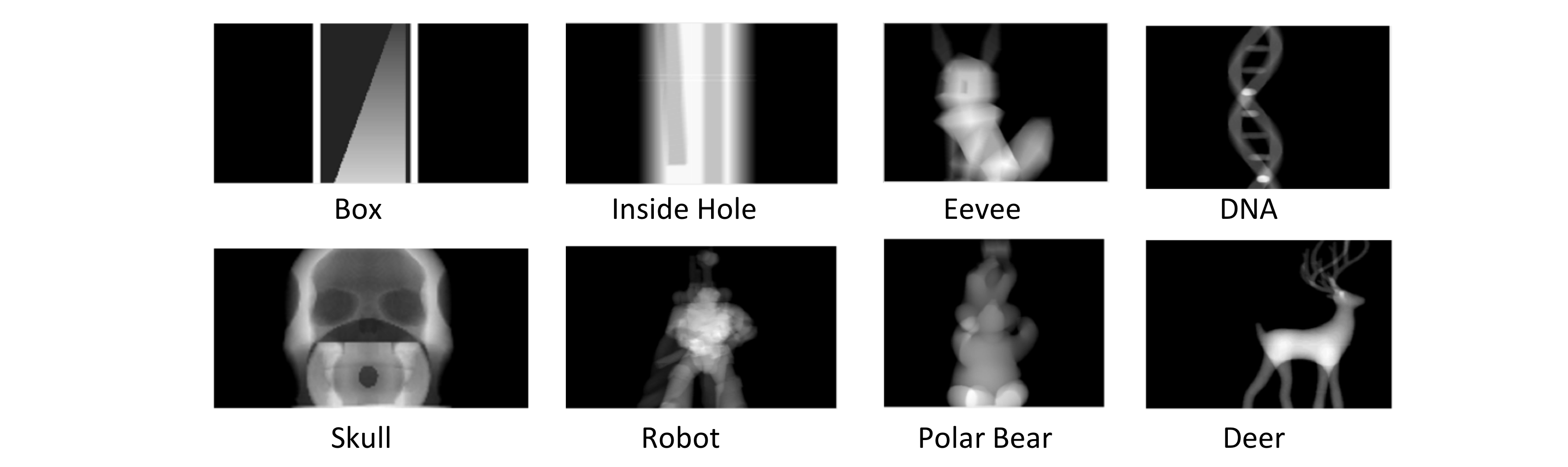}
    \caption{Illustration of the ground truth of eight objects in a projection angle used for training and testing (Box, Inside Hole, Eevee, DNA, Skull, Robot, Polar Bear, and Deer).}
    \label{fig:Dataset}
\end{figure}

As shown in Fig. \ref{fig:Dataset}, the THz dataset contains eight objects (Box, Inside Hole, Eevee, DNA, Skull, Robot, Polar Bear, and Deer), each including measured THz data along with its ground truth. The eight objects are designed due to their variety of geometry, which helps a THz DL-CT model reconstruct the unknown objects with more complex geometry. For example, the Robot and Deer objects possess high spatial frequency geometries, especially their body details and the high aspect ratio parts; The Box and Inside Hole objects, composed of a tiled cone inside the box, represent a hollowed, gradient structure inside a thick object. For each object, the dimension of the measured THz signal is $h \times 30 \times 288 \times 1000$ (2D slices, projections, width, time). Thus, stacking all the 2D slices in the objects, our total dataset has a dimension of $1349 \times 30 \times 288 \times 1000$. The dimension of the corresponding ground truth is $1349 \times 288 \times 288$ (2D slices, width, length).

\section{Model Implementation}

To develop a high-performance THz DL model for high-precision 3D image reconstruction, the model architecture contains four main features: low model complexity, high data usage efficiency, spatio-temporal data fusion, and image deblurring.

As building an end-to-end deep learning model for THz tomographic images requires a large memory utility (typically, more than 500 GByte per dataset), it exceeds the capability of the most currently available general-purpose graphics processing unit (GPU), impedes the development of data-driven THz CT modalities. To mitigate the memory burden, our THz DL-CT model maps the THz spatio-temporal data to a row in a sinogram of a cross-sectional image. THz 2D cross-sectional images can be reconstructed from the rows in the sinogram by the stacking and the inverse Radon transform. In the THz DL-CT model training phase, the ground truth cross-sectional images are pre-converted into the sinograms by Radon transform and the training targets are rows in sinograms of ground truth cross-sectional images. By this ground truth pre-conversion, the significant learning parameters in domain transformation between a sinogram and a cross-sectional image can be avoided. The input of the DL-CT model is the THz spatio-temporal data retrieved from decomposing by height and projection angles. This data decomposition can further relieve the computation power of learning the model compared to learning the mapping between the THz height-angle-spatio-temporal signal to the sinograms. By the learning strategy above, the THz DL-CT framework has at least two orders of magnitude fewer learning parameters compared to the end-to-end model.

In the spatio-temporal model architecture, to begin with, the THz DL-CT model learns the features from the THz temporal signals without considering the spatial correlation by designing the kernel size of 1 on the spatial axis. As the THz Gaussian beam interacts with multiple object voxels along the propagation path, the THz DL-CT model then addresses the object features in the local area and spatial correlation of neighbor signals. Based on the above learning approach, which separately takes the effect of temporal and spatial correlation into account, the THz DL-CT model can fully utilize all THz time-resolved data points and analyze the temporal profile correlations among the local regimes. This high spatio-temporal data usage efficiency enables better feature extraction capability than conventional THz CT methods.

\begin{figure}
    \centering
    \includegraphics[width=7cm]{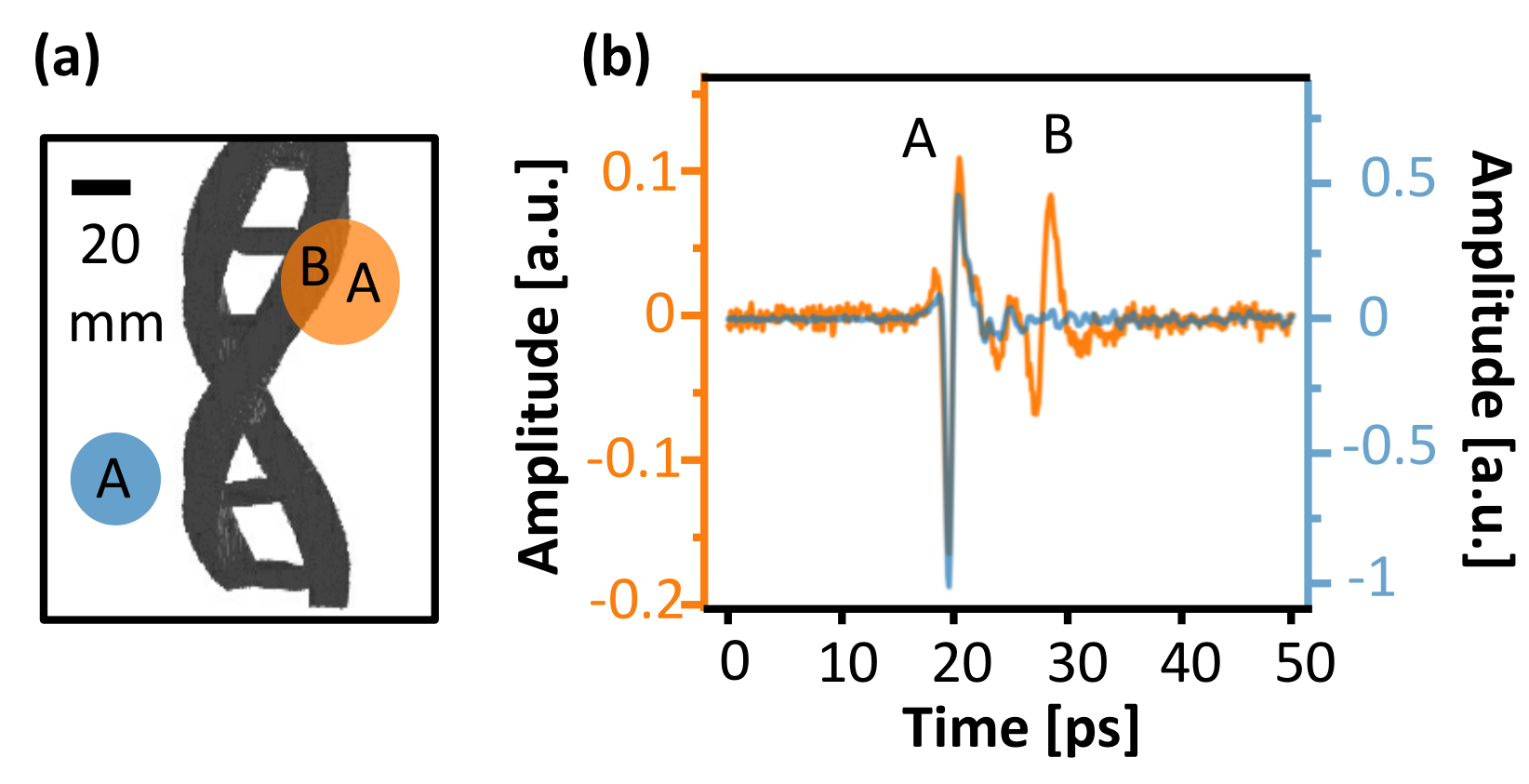}

    \caption{(a) The schematic diagram of the THz wave propagates through two object regions. In the orange region, THz waves pass through both "A" and "B" materials, whereas THz waves only pass "A" material in the blue region. (b) The measured THz signals in the blue and orange regions. The two dominant peaks correspond to the two materials, "A" and "B," respectively.}
    
    \label{fig:time_deviation}
\end{figure}

When the pulsed THz signal travels through a multi-material system, the dielectric response of each type of material introduces different time delay and signal distortion levels. This physical phenomenon combined with temporal signal features is included in the THz DL-CT model design for image deblurring at the material interface. For example, as shown in Fig.
\ref{fig:time_deviation}, the THz waves pass through material B with a larger refractive index, and higher absorption coefficients would cause more time delay and attenuated signal strength compared with material A. As the beam waist of propagated THz beam is much larger than the raster scanning step size, it causes the nearby-voxels information to mix in the measured THz time-resolved signals resulting in the imaging blurring problem. Considering the time delay and nearby-voxel information mixing, the THz DL-CT model first extracts the object information contained in the different peaks of THz time-resolved signals. Consequently, the THz DL-CT model pulls the spatio-temporal correlated features to unmix nearby-voxel information.

\begin{figure}
    \centering
    \includegraphics[width=\linewidth]{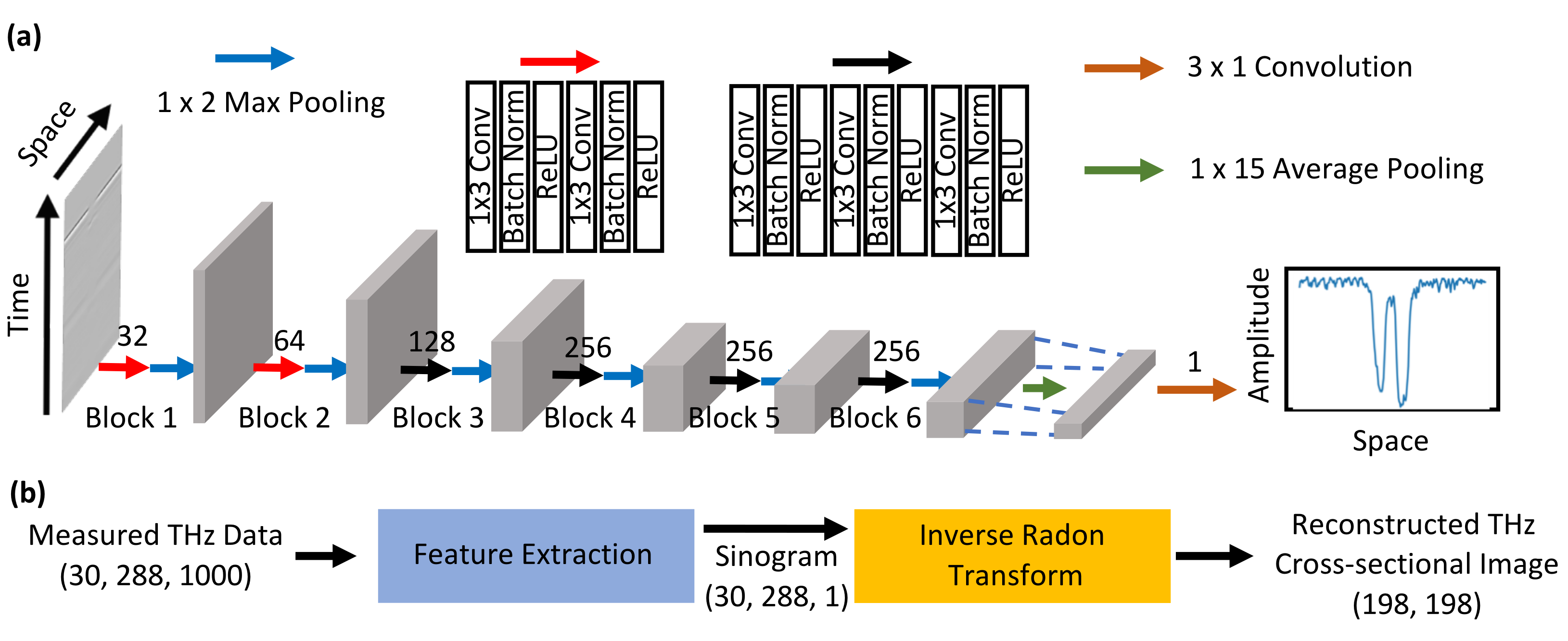}

    \caption{(a) The THz DL-CT model, where the input and output are THz spatio-temporal signal and a row of sinogram, respectively. The first six convolutional blocks (Block 1 to 6) and the last convolutional layer (brown arrow) are used to extract the temporal correlation and spatial correlation, respectively. The number above the arrows indicates the channel number. (b) The data flow of THz DL-CT framework. In the feature extraction block, the THz spatio-temporal signal of different projection angles is sequentially feedforwarded in the THz DL-CT model. The outputs of feature extraction are concatenated to a sinogram. By inverse Radon Transform, the THz cross-sectional image is reconstructed.}
    \label{fig:model_arch}
\end{figure}

The THz DL-CT model is based on the VGG16 network as the backbone to extract features from the spatio-temporal raw data, as shown in Fig. \ref{fig:model_arch}(a). According to the learning strategy mentioned above, time contraction without spatial contraction is implemented by the convolutional kernel size of $3\times1$ (time, space) until the last layer. The THz DL-CT model utilizes the convolutional kernel size of $1\times3$ to learn the spatial correlation in the last layer. Furthermore, to prevent the convergence instability caused by the high similarity between consecutive projections of measured THz time-domain traces in the cross-section, the order of projections feedforwarding into the model was randomly shuffled. In the THz cross-sectional image reconstruction phase, the data flow contains two main blocks, feature extraction, and Inverse Radon Transform, as shown in Fig. \ref{fig:model_arch}(b). In the feature extraction block, the THz spatio-temporal data with different projection angles consecutively feed into the THz DL-CT model to predict the THz CT sinogram; in the Inverse Radon Transform block, the predicted sinogram is transformed to the 2D cross-sectional images of the object. In the end, the THz 3D tomography image is visualized by stacking 2D cross-sectional images.

\FloatBarrier

\section{Results and Discussion}

\subsection{Comparison with Conventional and Machine Learning Methods}
\label{sec:comparison}

\begin{figure}
    \centering
    \includegraphics[width=\linewidth]{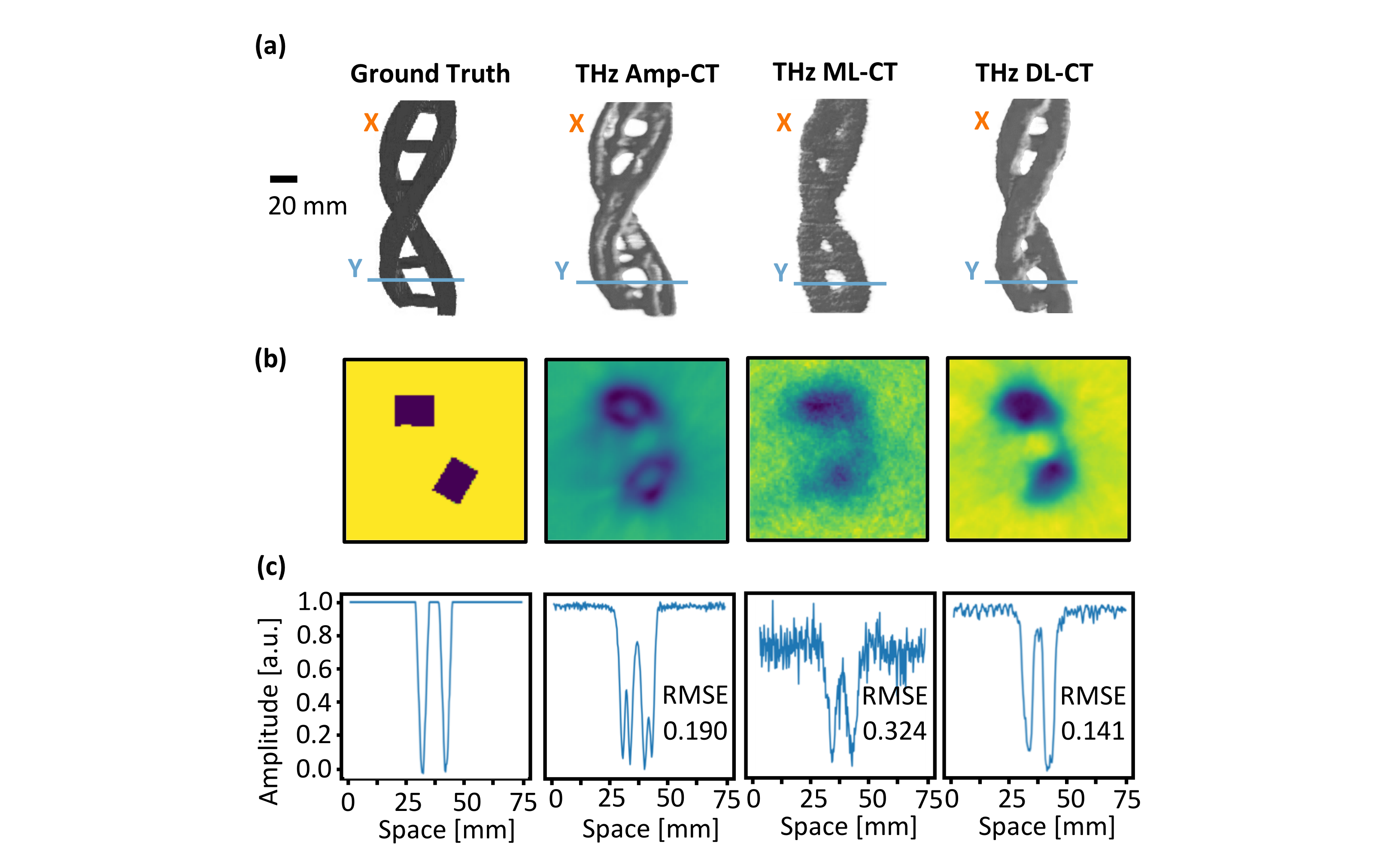}
    \caption{The comparison of ground truth, THz Amp-CT, THz ML-CT and THz DL-CT in (a) tomographic images, (b) cross-sectional images and (c) projected signals. The cross-sectional images are extracted from the "Y" cross section. The projected signals are extracted in the 108 degrees projection of the "Y" cross section, where the bottom of (b) is 0 degree with counter clockwise counting.}
    \label{fig:dna_comparison}
\end{figure}

To evaluate the performance of the THz DL-CT on THz 3D tomographic image reconstruction, THz maximum amplitude CT (THz Amp-CT) is selected as the compared method since it is one of the most robust and common methods in THz 3D tomographic imaging \cite{guerboukha2018toward}. In THz Amp-CT, the data flow of the reconstruction follows a similar approach as the THz DL-CT, except the feature extraction. The feature extraction of THz Amp-CT only selects the maximum value of each THz signal trace, which reflects the combination of object thickness and material absorption. Additionally, we design a THz machine learning CT (THz ML-CT) model with a polynomial regression model of degree 2 \cite{scikit-learn} to verify the efficacy of temporal and spatial contraction in the THz DL-CT model. The THz ML-CT model takes both the THz time-resolved signals and the element-wise square of the THz time-resolved signals to construct a linear polynomial regression mapping to predict a sinogram of the object under test. To this extent, THz ML-CT is designed not addressing on the time deviated issue nor the issue of nearby-voxel information mixing. A 3D-printed DNA-shaped object is then performed as the test object. Its double-helix geometry contains many geometric features, including separated blocks, twisted shapes, periodic separation, which is well applicable to evaluate the image reconstruction performance. Moreover, the single helix and the gap between double helix can test the imaging artifact level and the imaging resolution, respectively. To analyze the performance of the three THz CT methods, the 3D tomographic image is further decomposed into cross-sectional images and rows in sinogram of cross-sectional images (as shown in Fig. \ref{fig:dna_comparison}).

Fig. \ref{fig:dna_comparison}(a) shows the comparison of the reconstructed THz CT images among THz Amp-CT, THz ML-CT, and THz DL-CT. While THz waves travel through the 3D-printed DNA-shaped object, the energy of the pulsed THz signal is distributed into a broader time window. Due to this energy relocation, THz Amp-CT filters only preserve limited object information causing the false hollow artifact in the single helix region (i.e., orange "X"). Compared with THz Amp-CT, THz ML-CT includes the whole THz time-resolved signal and its energy distribution in a 100 ps timeframe. With the additional energy information and temporal profile features, the THz ML-CT can analyze more object information, including material absorption levels and object thickness distribution; thus, THz ML-CT decreases the artifact levels and enhances the image structural similarity, especially inside the object. However, the additional energy information also elevates noise level, resulting in higher surface roughness in the reconstructed 3D image. In THz DL-CT, the object surface is much smoother than THz ML-CT since the THz DL-CT model has already suppressed the noise level during the learning process \cite{image_denoise}. With the combination of the denoising feature and additional spatio-temporal information, THz DL-CT offers the best image quality in accuracy, artifact level, and shape similarity. In the cross-sectional image, as shown in Fig. \ref{fig:dna_comparison}(b) (i.e., the blue "Y" layer in Fig.
\ref{fig:dna_comparison}(a)), THz Amp-CT present worse image quality mainly caused by energy distribution in time. It introduces the two hollowed artificial defects inside the two single-helix shapes and low image contrast because the maximum THz signal values are no longer linearly proportional to the ground truth. The THz ML-CT demonstrates decent imaging contrast since the THz ML-CT model can shift and scale the input to address the offset between the input and ground truth. However, the additional energy information introduces pepper noise over the image due to the Gaussian distribution of the measurement system noise. In the THz DL-CT model, the imaging contrast is greatly enhanced due to the convolutional architecture of the model \cite{kuang2019single}. In addition, since the input of the THz DL-CT model does not include the energy information of THz time-resolved signals, the THz DL-CT model is not required to address the additional noise issue in the energy information of THz time-resolved signals. Therefore, the THz DL-CT delivers superior imaging quality in noise level and object geometry estimation and can adopt a material system with more complex object geometry. To evaluate the three models quantitatively, we analyze projected signals in all predicted sinograms of different cross-sectional object images. In Fig. \ref{fig:dna_comparison}(c), it shows the projected signal of 108 degrees in the blue "Y" cross-section (0 degree is at the bottom and degrees count with counter clockwise), which presents the reconstructed signal strength, quality, and gap region along the concentric line of the double helix. As THz Amp-CT only extracts the maximum value of the time-resolved signal without time registration, the two sharp peaks inside the double valley show the artificial vacancies of the object, resulting in inferior root mean square error (RMSE) compared to THz DL-CT. THz ML-CT demonstrates decent object structures; however, the additional system noise significantly reduces signal quality to a 0.324 RMSE. Compared with THz Amp-CT and THz ML-CT, THz DL-CT improves at least 25.8\% in RMSE and demonstrates excellent object estimation with much less noise. Since the available THz-TDS imaging dataset is limited, cross-validation training is applied to our dataset to investigate the imaging efficacy of THz DL-CT on different object geometry.

As shown in Table. \ref{tab:t1}, THz DL-CT outperforms THz Amp-CT and THz ML-CT in every tested object. By learning the temporal and spatial correlation in different layers in the THz DL-CT model, THz DL-CT reconstructs the precise THz 3D tomographic image with low noise introduced and excellent object geometry estimation and shows at least 50.2\%, and 52.6 superior imaging efficacy in terms of RMSE and structural similarity index measure (SSIM) respectively, compared with THz Amp-CT and THz ML-CT. It demonstrates that the superior reconstructed image quality of THz DL-CT is not just come from the data-driven approach. Most importantly, the THz spatio-temporal data feature extraction significantly improves the accuracy and structure of the reconstructed images. 

\FloatBarrier

\begin{table}
  \centering
  \caption{Quantitative cross-sections comparison (RMSE and SSIM) of THz DL-CT, THz Amp-CT, and THz ML-CT on \textbf{DNA}, \textbf{Box}, \textbf{Deer}, \textbf{Eevee}, \textbf{Inside Hole}, \textbf{Polar Bear}, \textbf{Robot}, and \textbf{Skull}. ($\downarrow$: lower is better) ($\uparrow$: higher is better)}
  \includegraphics[width=1\linewidth]{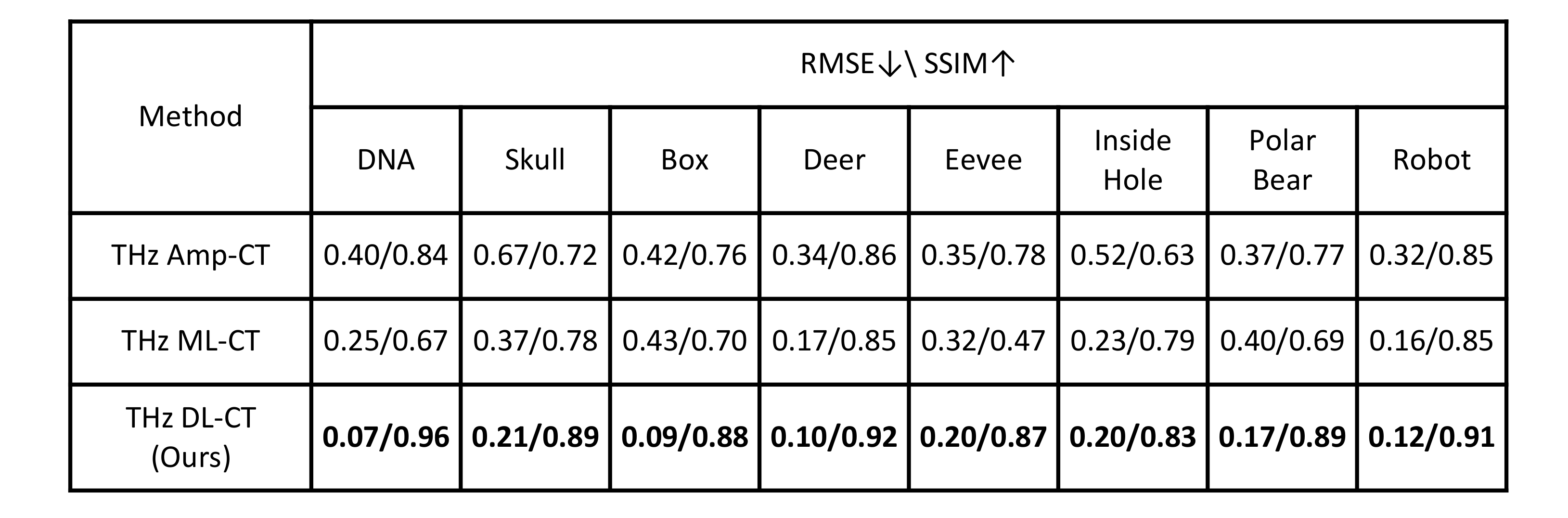}
  \label{tab:t1}
\end{table}

\FloatBarrier

\subsection{Kernel Placement Discussion}


The THz DL-CT utilizes the temporal profiles and the nearby-voxel information mixing among the THz spatio-temporal signal by learning the temporal and spatial features in the different blocks. Addressing temporal information, the THz DL-CT model contains the global average layer and multiple blocks composed by $1\times3$ kernels, batch normalization, and activation function to extract the object information located in different temporal regions of the time-resolved signals. With the preliminary learned temporal features, the THz DL-CT is designed to allocate a $3\times1$ kernel block for the nearby-voxel information mixing. To investigate the impact of spatial kernel placement order on the processing scope of spatio-temporal information, we then evaluate the three THz DL-CT model architectures with respective spatial kernel placements in the first layer of first, fourth and sixth (last) convolutional blocks, denoted as 1-1 THz DL-CT model, 4-1 THz DL-CT model and THz DL-CT model (as shown in Fig. \ref{fig:model_arch}).

\begin{figure}
    \centering
    \includegraphics[width=7cm]{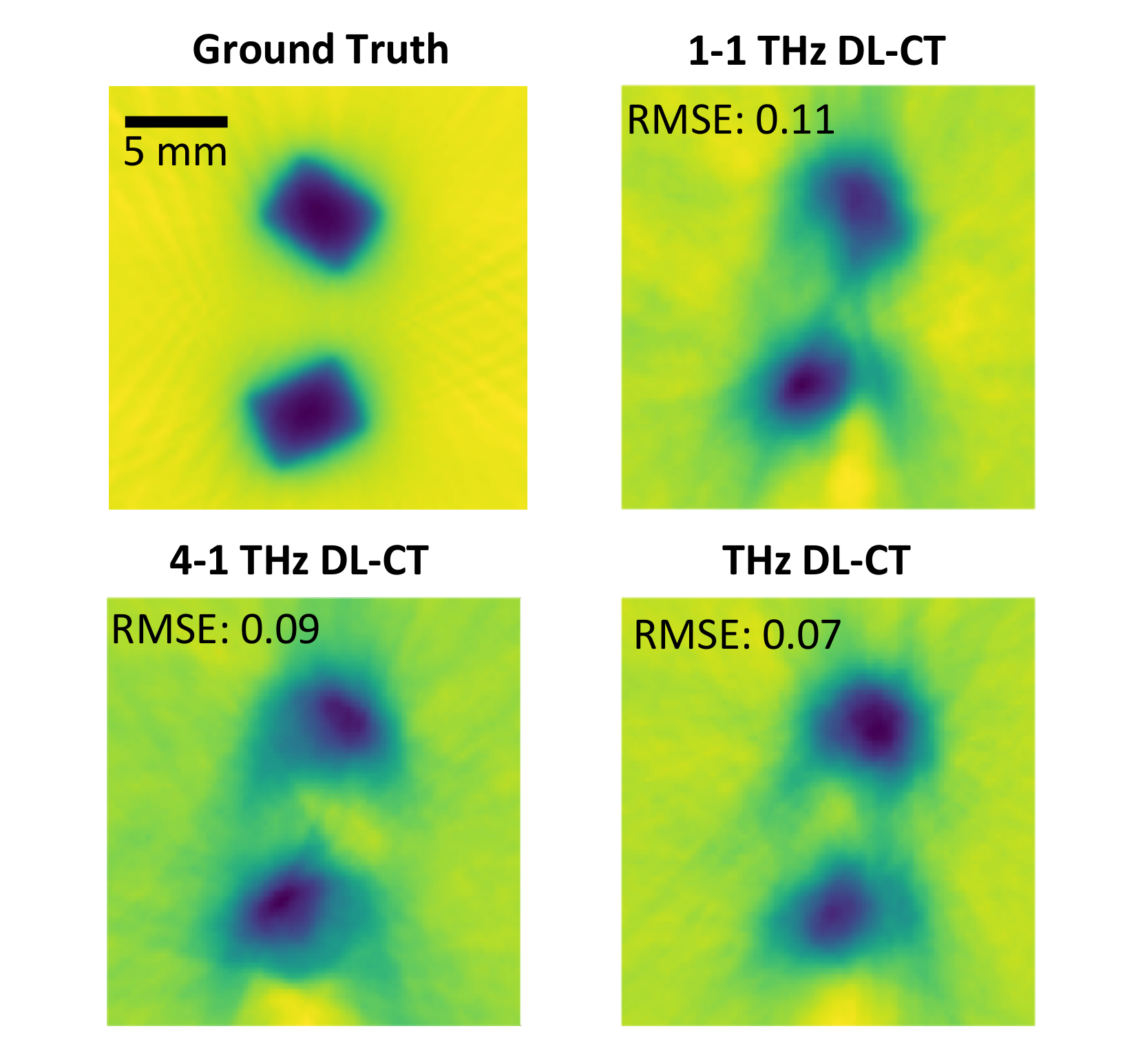}
    \caption{The ground truth and the comparison of THz cross-sectional images by three variants of the THz DL-CT models with spatial kernel placements in the first layer of first (1-1 THz DL-CT), fourth (4-1 THz DL-CT model) and last (THz DL-CT model) convolutional blocks.}
    \label{fig:kernel_placement_image}
\end{figure}

In the 1-1 THz DL-CT model, the receptive field of the spatial kernel covers the 0.5 ps among the 100 ps THz time-resolved signal. Since this sub-ps temporal interval contains the temporal features corresponding to the sub-mm range in the spatial domain, the 1-1 THz DL-CT model can just process localized object information in the dimension of approximately a single wavelength. The 4-1 THz DL-CT model exhibits a 9.2 ps temporal receptive field in spatial contraction to simultaneously capture the geometric details in local regions and mm-scale object features in the broader scope. However, this 9.2 ps receptive field can still cover limited numbers of the pulsed THz signal echo bounced back from object boundaries. As the topological details of the object are carried in multi-bounce THz signal, spreading to tens of ps time frame while the object is in several cm sizes, it is challenging to accurately resolve the macroscopic body of the object with such receptive field scope. By placing the spatial kernel in the last convolution block, the THz DL-CT model has a wider spatio-temporal field of view in nature to extract the object geometry in localized details, regional features, and object topology on the macroscopic scale. As shown in Fig. \ref{fig:kernel_placement_image}, the 1-1 THz DL-CT model presents a more blurry image in the boundary of the double helix and lower image contrast. This is because the sub-ps temporal window of 1-1 THz DL-CT model constrains the processing geometric scope down to sub-mm scale, which leads to the low resolution in the sinogram of the THz cross-sectional image. Compared to the 4-1 THz DL-CT model, the THz DL-CT model can utilize the spatio-temporal information along with the multiple bouncing THz pathways; thus, the THz DL-CT model delivers a much sharper boundary of the double helix and precise object location. With a wider spatio-temporal field of view, utilizing the dataset from microscopically to macroscopically, the THz DL-CT model achieves a RMSE of 0.07, which improves 36.4 \% compared to the 1-1 THz DL-CT model.

Unlike conventional physics-model-based THz CT methods, the THz DL-CT model does not need prior knowledge of the object information like material content or physical models to reconstruct object 3D geometry. In practice, data-driven approaches can achieve a 100\% data use rate, preserving combinations of light-matter interactions, including geometric-dependant absorption, reflection, refraction, and scattering phenomena. This 100\% data usage rate is several orders of magnitude higher than physics-model-based approaches and no more constrained by specific physical models to extract object features. As there is no physical model integrated into the THz DL-CT model, it is crucial to understand how our data-driven model addresses spatio-temporal data. From unveiling the black box of deep learning model point of view, the relationship between extensively used data portions and the physical phenomenon could be an exciting research field to explore. To this extent, we use the saliency map \cite{saliency_map} to visualize the weight of the three THz DL-CT model variants and further evaluate their feature extraction approaches in the THz spatio-temporal signal. In the saliency map, greater values of pixels indicate that models pay more attention to those pixels for the prediction. More specifically, the pixels with more attention from the model are more likely to influence the prediction decision.

\begin{figure}
    \centering
    \includegraphics[width=7cm]{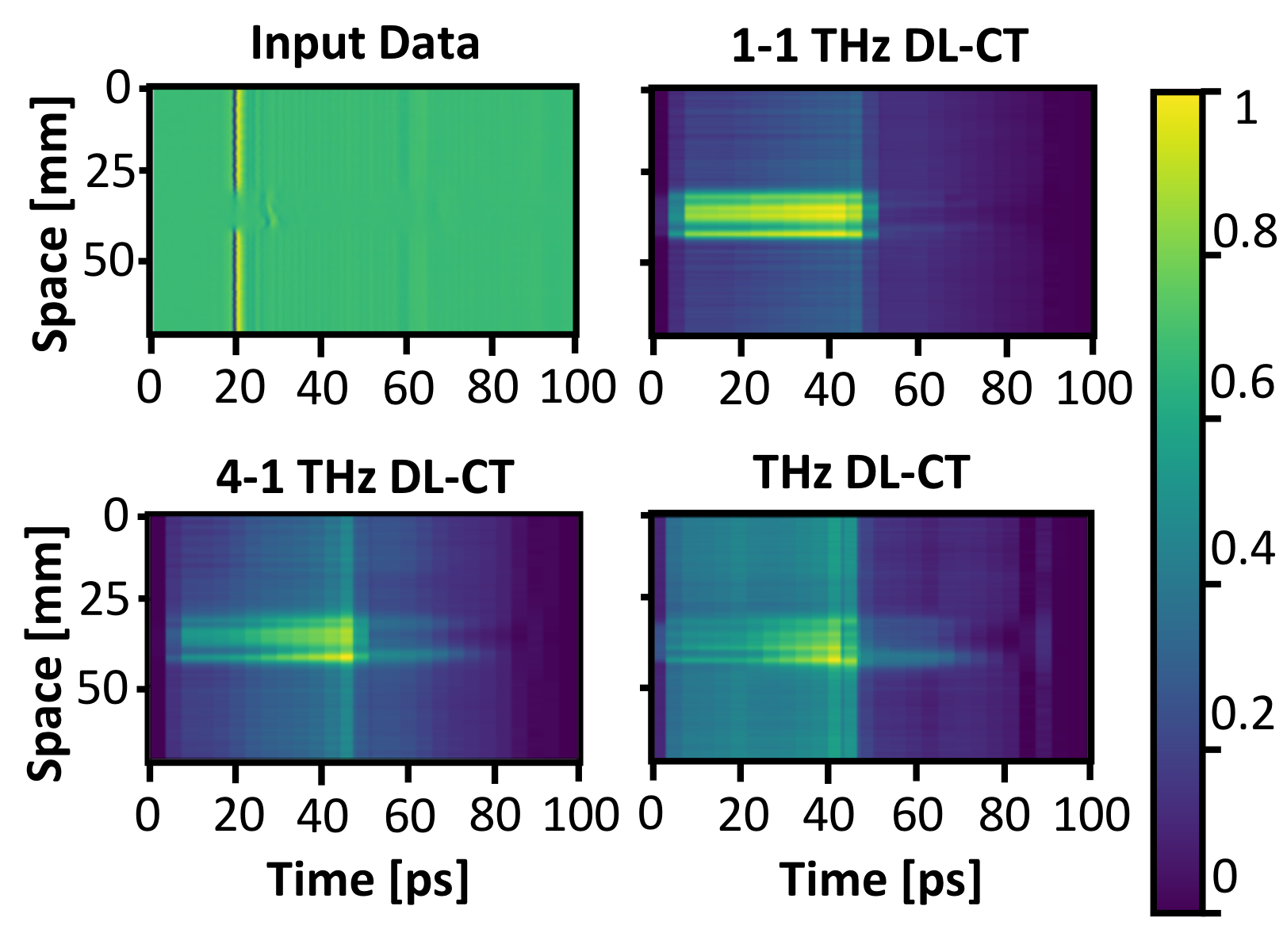}
    \caption{The saliency map of three variants of the THz DL-CT models with different spatial kernel placement order. The input data is the THz spatio-temporal signal of 108 degrees projection in the cross-section in Fig. \ref{fig:kernel_placement_image}. }
    \label{fig:kernel_placement_saliency}
\end{figure}

The THz spatio-temporal signal corresponding to the same 108 degrees as the above section is selected as input because of the multiple signal characteristics – time delayed, attenuated signal by material absorption, and extended multiple bounced THz signals (as shown in Fig. \ref{fig:kernel_placement_saliency}). The time delay signal presented in the spatio-temporal region of 30 -- 40 mm and 20 -- 50 ps encodes the primary material absorption and thickness information of the tested object. The first bounced THz signal carries thickness, and object boundary information is located at 60 -- 70 ps. As expected, the 1-1 THz DL-CT model put the most attention temporal signal close to the object area (i.e., 30 to 40 mm) to estimate the object thickness. In contrast to the 1-1 THz DL-CT model, the 4-1 THz DL-CT model can compare the air and object area signals to extract object information and utilize the signal in the area of 40 -- 50 ps, which also contains certain background noise. Allocating the spatial kernel to the last layer, the THz DL-CT model can take the multiple bounced THz signal echoes into account to better profile the object geometry.

\FloatBarrier

\subsection{Application}
In the sections above, the THz DL-CT has demonstrated the superior imaging performance of RMSE and SSIM on the 3D objects compared with conventional and machine learning methods. The THz DL-CT can reconstruct the object geometry by efficiently using the THz spatio-temporal signal, which contains rich features, such as energy absorption, time delay, and boundary information. Additionally, the THz DL-CT does not require prior object information like material contents. With those advantages of 100\% data usage and no requirement of any prior object information, the THz DL-CT framework could be suitable for more complex material systems. More specifically, the THz DL-CT framework trained on 3D-printed material objects can be generalized to reconstruct different multi-material object systems, which do not correlate with the material of the training objects.  

\begin{figure}
    \centering
    \includegraphics[width=7cm]{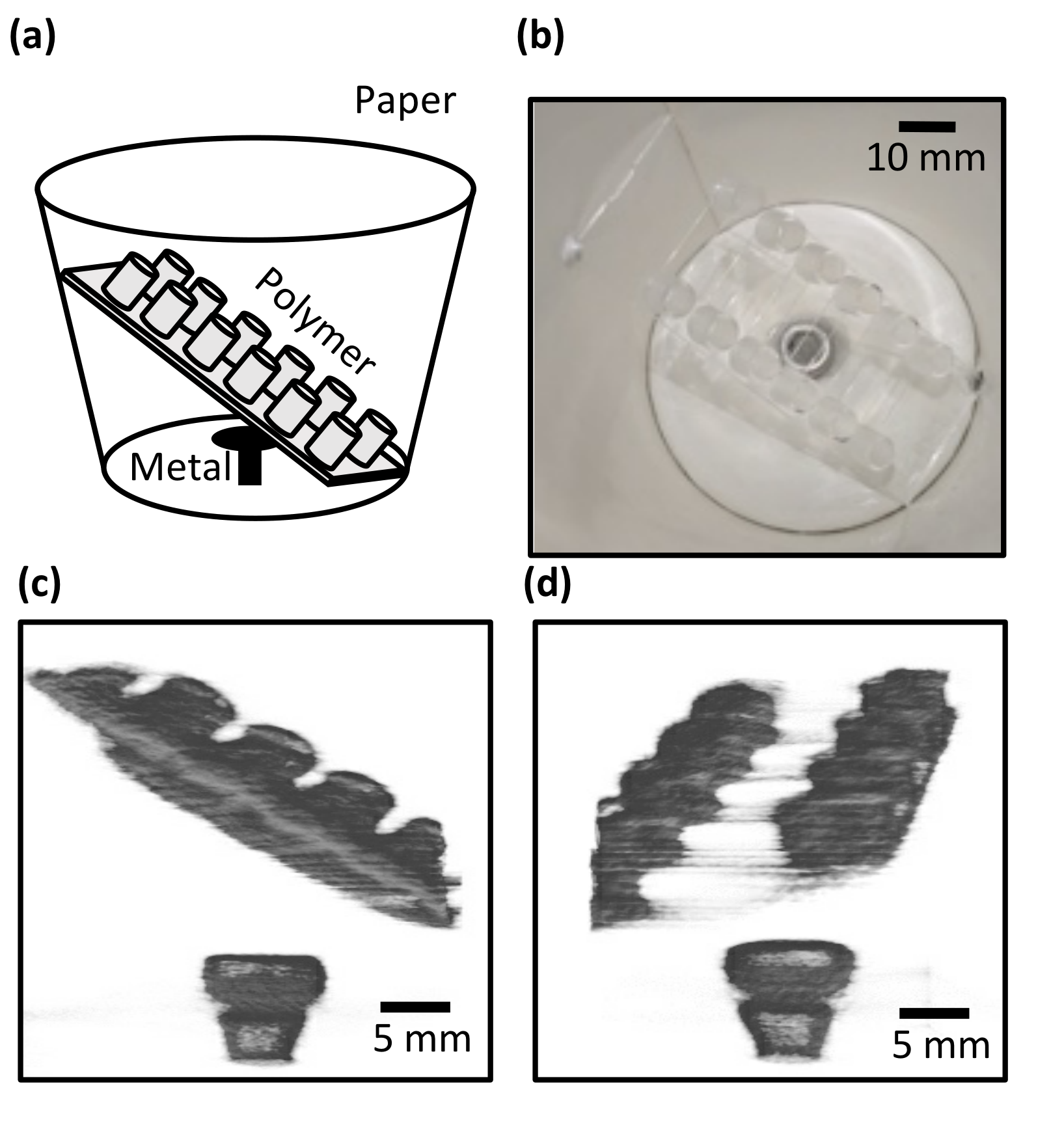}
    \caption{(a) The illustration and (b) optical image of the measured multi-material system containing a paper cup, an polymer platform and a metal screw. (c)(d) The THz tomographic images of the multi-material system reconstructed from the THz DL-CT model.}
    \label{fig:multi_material}
\end{figure}

To demonstrate this concept, we design a multi-material system composed of a paper cup, a polymer platform, and a metal screw representing the varying extents of THz dielectric responses. For example, the paper and polymer are semi-transparent in the THz range; metal represents its highly reflective nature in the THz range. As shown in Fig. \ref{fig:multi_material}(a), the polymer platform with multiple cylinder pipes is placed inside the paper cup, and the metal screw beneath the polymer object is attached to the bottom of the paper cup. In this pilot study, we use the same THz DL-CT model as the one in Section \ref{sec:comparison}. The scanning and reconstruction configurations remain the same as described in the previous sections. As shown in Fig. \ref{fig:multi_material}(c)(d), the THz DL-CT presents its capability of reconstructing the periodic cylinder shape of polymer in the multi-material system with the sub-millimeter spatial resolution. Although polymer and metal are placed in the optically opaque paper cup, THz DL-CT can still clearly show the gap between polymer and metal. It should be noted that the region of the paper cup in cross-sectional images is cropped after the reconstruction for visualization purposes.

\FloatBarrier

\section{Conclusion}
In this paper, we propose a THz DL-CT framework utilizing the THz spatio-temporal signal to reconstruct the THz tomographic images without any prior object information. This THz DL-CT framework is specifically designed based on two objectives: reduction of learning parameters and the physics-inspired learning strategy. Instead of building an end-to-end model, the THz DL-CT model takes rows in the sinogram of the ground truth cross-sectional images as training targets, which can tremendously reduce training parameters in the order of one billion. Furthermore, dividing the THz spatio-temporal signals of different angles and layers as input of the THz DL-CT model can further reduce three orders of magnitudes of learning parameters compared to launch THz height-angle-spatio-temporal signals as the input directly. By the two data preparation strategies of reducing learning parameters, the THz DL-CT framework can be trained on the general-purpose GPU.

The THz DL-CT model extracts the object information located in the different regions of the THz spatio-temporal signal by temporal and spatial contraction. With the use of temporal convolutional blocks, the THz DL-CT model can extract the features from the time delay signal and the signal of multi-bounce light paths; Followed by the temporal convolutional blocks, the spatial convolutional blocks can address the nearby-voxel mixing information issue resulting in higher image accuracy and sharper boundary edges of the reconstructed THz images. Furthermore, the contrast of the reconstructed THz cross-sectional image is improved by the architecture of the convolutional neural networks. With those designs and the learning strategy, the THz DL-CT model demonstrates the THz cross-sectional images with higher image contrast, less artifacts, and superior imaging efficacy of SSIM and RMSE. Here, the THz DL-CT model achieves a 65.8\% and a 52.6\% in RMSE and SSIM better than the conventional THz model-based method, respectively. Compared to the conventional physics-model-based THz CT methods, the THz DL-CT model utilizes the 100\% temporal signal without both prior knowledge and the use of physical models. To understand how the THz DL-CT model extracts the object features, which is different from model-based approaches, we then visualize the weight attentions on THz spatio-temporal signal of THz DL-CT through the saliency map. By evaluating the saliency map of the THz DL-CT model (space-time-time learning strategy) and its variants (time-space-time learning strategy), the THz DL-CT model has a wider spatio-temporal field of view to overview the time delay signals, distorted signals by material absorption dispersion, and the echo signals in the different temporal regions. By utilizing this overview capability, the THz DL-CT model has demonstrated its superior performance to extract the object information and delivers the excellent imaging efficacy of RMSE.

Since the THz DL-CT model does not require any prior information of the measured object, such as geometry and material, the pretrained THz DL-CT model on the single-material objects can deliver the acceptable imaging efficacy on the multi-material systems without any fine-tunning. To demonstrate this concept, a multi-material system composed of paper, polymer, and metal representing the varying extents to THz light-matter interaction is designed and reconstructed by the pretrained THz DL-CT model on single-material objects composed by HIPS. The THz DL-CT model can deliver the sub-millimeters spatial resolution and reconstruct the details of the multi-material system, such as the metal screw, the polymer cylinders, and the gaps between them. With this feature, the cost of deploying a universal THz DL-CT framework on a great variety of multi-material systems can be significantly decreased without trading off the imaging performance. Functional imaging is a great extent to this work since most of the nonpolar substances have the unique ``fingerprints'' in THz bands \cite{zhong2006identification, lin2021hyperion, davies2008terahertz}. Combining the additional model handling the material identification to the THz DL-CT framework, not only the object behavior but also the material identification can be achieved. Furthermore, the correlation among spatio-temporal data in the consecutive projection angles and cross-sections can also be considered to deliver the more precise object geometry estimation. To reconstruct a superior 3D tomographic image, the regularization on cross-sectional images can also be used to encourage the smooth reconstructed surface or the sharp edge 
\cite{velikina2007limited, hamalainen2013sparse}.



\section{Data Availability Statement}
Data underlying the results presented in this paper are not publicly available at this time but may be obtained from the authors upon reasonable request.

\bibliographystyle{unsrt}  
\bibliography{references}

\end{document}